\documentclass[reprint,amsmath,amssymb,aps,prl]{revtex4-2}
\usepackage{graphicx}
\usepackage{dcolumn}
\usepackage{bm}
\usepackage{xcolor}

\begin{document}

\title{{Dynamic pressure enhancement upon disk impact on a boiling liquid\\}}

\author{Yee Li (Ellis) Fan$^1$}
\email{Contact author: ellisfan179@gmail.com}
\author{Bernardo Palacios Muñiz$^1$}
\author{Nayoung Kim$^1$} 
\author{Devaraj van der Meer$^1$}
\email{Contact author: d.vandermeer@utwente.nl}

\affiliation{%
$^1$Physics of Fluids Group and Max Planck Center Twente for Complex Fluid Dynamics,
MESA+ Institute and J. M. Burgers Centre for Fluid Dynamics, University of Twente,
P.O. Box 217, 7500AE Enschede, The Netherlands}%

\date{\today}

\begin{abstract}
We experimentally investigate the impact of a flat, horizontal disk onto a boiling liquid, i.e., a liquid in thermal equilibrium with its vapor phase. We observe exceptionally high impact pressures deviating strongly from the inertial scaling found for impact in a non-condensable environment, coinciding with the rapid collapse of the vapor pocket entrapped below the disk. We explain our findings, which are relevant for the safe transportation of cryogenic fuels, as a result of vapor condensation, leading to accelerated vapor pocket contraction at high impact velocity and low vapor density.
\end{abstract}


\maketitle

In any solid-liquid impact, whether one deals with a flat \cite{ermanyuk2011experimental,peters2013splash,jain2021air} or curved \cite{marston2011bubble,hicks2012air,carrat2023air} solid body impacting onto a liquid pool, or with the impact of liquid onto a solid surface \cite{bagnold1939interim,bouwhuis2012maximal,tran2013air,josserand2016drop}, air entrapment is a ubiquitous and unavoidable phenomenon, known to provide air-cushioning that mitigates the maximum impact load exerted on the solid body \cite{verhagen1967impact,jain2021air,kim2021water,fan2024air}. It is widely agreed upon that this occurs due to the pressure build-up in the intervening air layer that deforms the liquid surface before impact, resulting in the entrapment of air by the pinch-off of the enclosed air pocket upon impact \cite{lee2012does,bouwhuis2015initial,hendrix2016universal}. While the entrapment of non-condensable air effectively cushions solid-liquid impact, this may no longer hold during boiling liquid impact, where the liquid is in thermal equilibrium with its vapor phase and a condensable vapor pocket is entrapped. With a boiling liquid, a minute increase of pressure --as happens during an impact event-- can disturb the equilibrium and may induce a phase change. 

Understanding the dynamics of boiling liquid impact is crucial for the safe transportation of cryogenic fuels, such as liquified natural gas (LNG) or liquid hydrogen (LH2) which are pivotal to the energy transition. In addition, this Letter is relevant to the development of LH2-fueled air- and spacecraft, water-hammer phenomena in pipe flow and the fundamental understanding of the role of phase change in multiphase flows. Cryogenic fuels are usually transported in containment tanks at atmospheric or increased pressure, such that they are always in a boiling state, i.e., close to thermodynamic equilibrium with their vapor phase. Slamming of the liquid on the containment wall, commonly known as sloshing, may induce a sudden high impact load during transportation \cite{dias2018slamming}.  Experimental studies on boiling liquid impact, i.e. solid-liquid impact in a liquid-vapor system are limited and are mostly at the larger scale of sloshing wave impact \cite{maillard2009influence,lee2021experimental}. Theoretical analysis of the pre-impact stage of LNG-solid impact that assumed a viscous vapor lubrication film suggested that phase change may reduce the pressure buildup in the gas layer, causing less deceleration of the liquid prior to impact \cite{hicks2018lng}. With this, it was postulated that a higher impact pressure may be induced upon impact as less momentum is dissipated before impact. 

In this Letter, we experimentally demonstrate that condensation of the entrapped vapor pocket can lead to an increase in excess of an order of magnitude of the local impact pressure on the impacting solid body and, using scaling arguments, in addition explore the conditions under which such an increase may occur. In contrast to \cite{hicks2018lng}, we highlight the inertial, condensation-induced collapse of the vapor layer and its effect on enhancing the impact pressure. Here, it is good to note that the reduced Reynolds number ($Re_\text{lub} \sim \rho_\text{v,0}U_0h_0/\mu_\text{v,0}$) is much greater than 1 in our Letter \cite{jain2021KH}, and hence, the vapor flow is inviscid.

\begin{figure*}[t!]
\centering
    \includegraphics[trim=0cm 0cm 0cm 0cm,width=.62\textwidth]{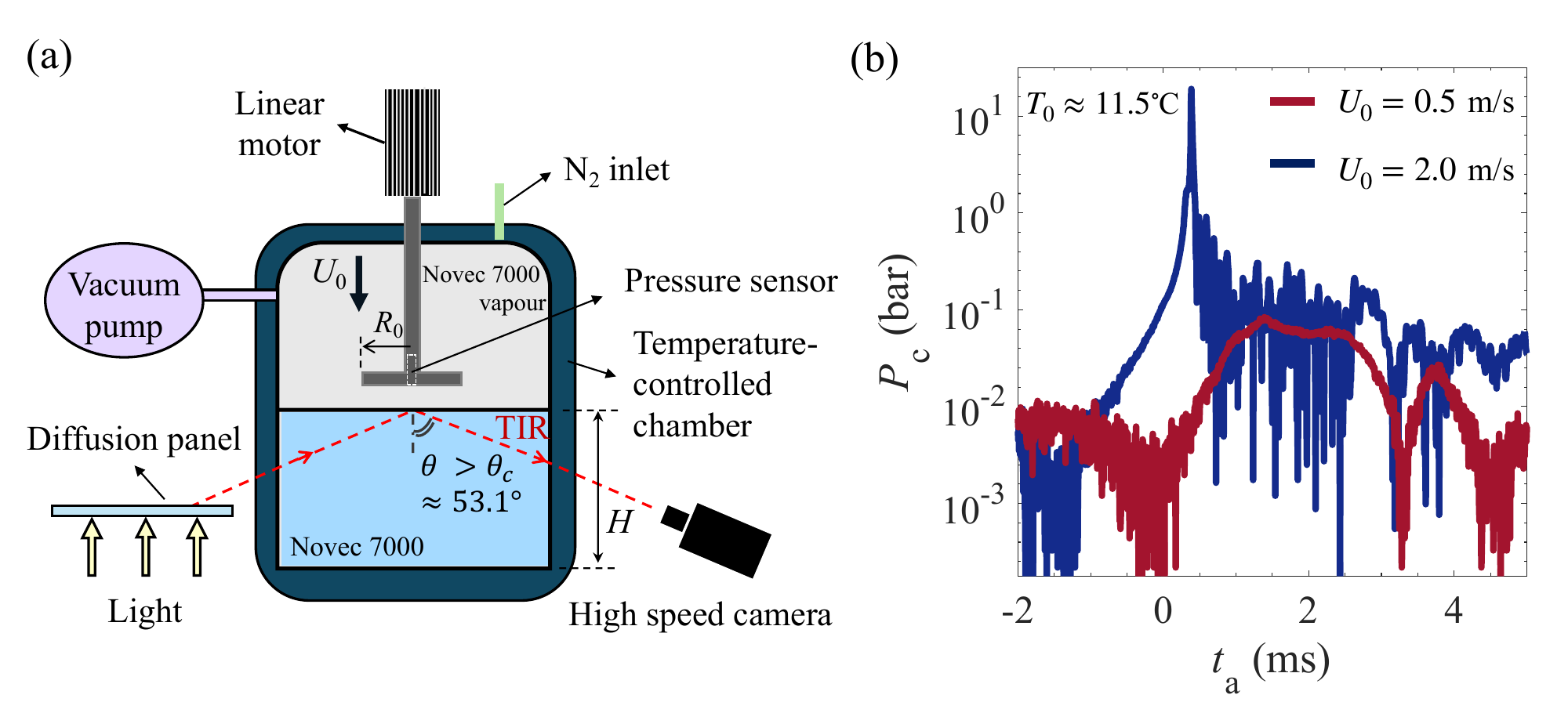} \hfill
    \includegraphics[trim=1cm 0cm 0cm 0cm,width=.36\textwidth]{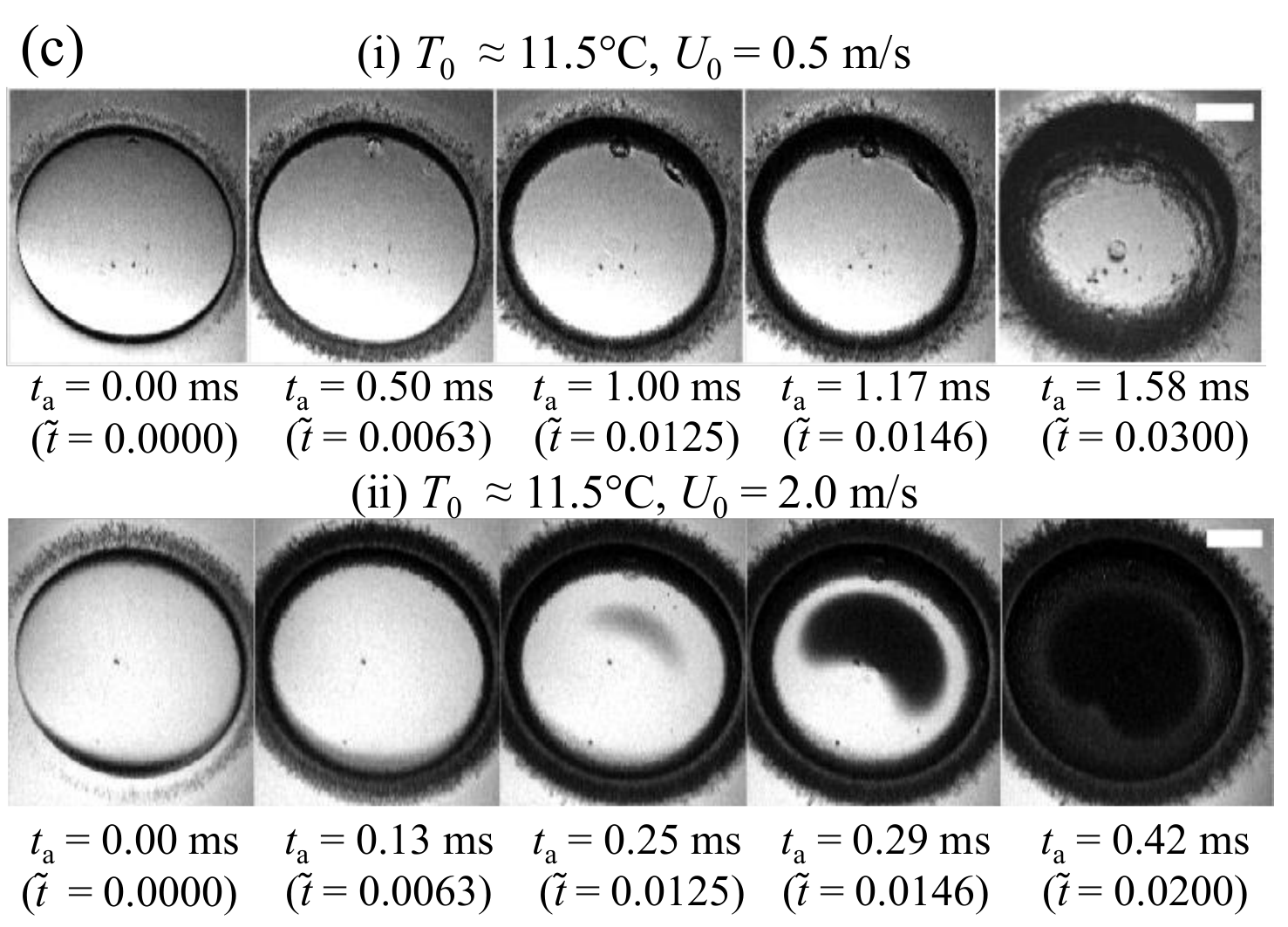} 
\caption{\label{fig:fig1} (a) Schematic of the experimental setup. (b) Time evolution of the measured pressure $P_c$ at the center of the disk, on a logarithmic scale, for two different impact speeds $U_0 = 0.5$ and $2.0$ m/s at an ambient temperature $T_0 \approx 11.5$ $^\circ$C. Here, $t_\text{a} = 0$ corresponds to the time at which the pressure signal at a reference pressure sensor near the disk edge rises above 10$\%$ of its maximum value. (c) Five re-aspected snapshots from the evolution of the entrapped vapor pocket under the impacting disk shown in (b). Note that, $t_\text{a} = 0$ coincides with the impact time, i.e., the time at which the disk is observed to make initial contact with the liquid surface from the high-speed recordings and $\tilde{t} = U_0t_\text{a}/R_0$ is the dimensionless time. (i) Gradual retraction of the liquid-vapor contact line from the disk edge at $T_0 \approx 11.5$ $^{\circ}$C and $U_0 = 0.5$ m/s. The entrapped vapor pocket is punctured around the disk center only at a very late stage (see Supplementary video 1 \cite{supplemental}). (ii) Rapid collapse of the entrapped vapor pocket at $T_0 \approx 11.5^{\circ}$C and $U_0$ = 2.0 m/s, where the entrapped vapor condenses into liquid around the disk center, resulting in the dark shaded wetted inner region of the disk that expands rapidly with time (see also in Supplementary video 2 \cite{supplemental}). [Scale bar = 20 mm]
}
\end{figure*}

To study the dynamics of boiling liquid impact, we build an experimental setup consisting of a 30 $\times$ 30 $\times$ 30 cm sealed chamber with circulating channels on its walls as depicted in Fig. \ref{fig:fig1}a. This allows the control of the ambient temperature $T_{0}$ within the chamber by circulating water with a chiller. To achieve the boiling state, in which the liquid is in thermal equilibrium with merely its own vapor, the chamber is first evacuated with a vacuum pump and flushed with nitrogen gas $(\text{N}_2)$ twice to eliminate as much moist air as possible from the chamber. Then, the working fluid, Novec 7000 (C$_4$F$_7$OH$_3$) is released into the evacuated chamber from a sealed reservoir. Novec 7000 is used owing to its low saturation vapor pressure at room temperature and its low boiling point. Its properties have been characterized to a sufficient extent in the literature \cite{3MNovec,widiatmo2001equations,ohta2001liquid,rausch2015density,perkins2022measurement,aminian2022ideal} and are summarized in the Supplemental Material~\cite{supplemental}. Liquid Novec 7000 vaporizes immediately upon entering the evacuated chamber and the pressure in the chamber $p_{\text{v,0}}$ increases towards its saturation pressure at the desired ambient temperature $T_0$. Subsequently, liquid Novec 7000 will start to fill the chamber. 

A circular horizontal flat disk of radius $R_0 = $ 40 mm and made of steel (RVS 316) is mounted to a linear motor that controls the impact velocity $U_0$ of the disk through a rod and is impacted vertically upon the liquid bath with its impacting surface parallel to the free liquid surface. The impact velocity $U_0$ is varied from 0.5 m/s to 2.0 m/s at 0.25 m/s intervals. In addition, five different ambient temperatures ranging from $T_0$ $\approx$ 11.5 $-$ 24.5 $^{\circ}$C, are tested in the experiment with the corresponding saturation pressure ranging from $p_{\text{v,0}}$ $\approx$ 420 $-$ 740 mbar. One thing to note is that the saturation pressure reached in our system is always higher than the reference saturation pressure reported in \cite{widiatmo2001equations} by approximately 50 mbar, which we believe is largely due to air dissolved in the liquid. However, due to the large density difference, air is expected to be stabilized above the Novec 7000 vapor and has minimal influence on the vapor entrapment at the liquid-gas interface.

A Kistler pressure sensor (Type 601C) is flush mounted at the disk center to measure the impact pressure at an acquisition rate of 200 kHz. By illuminating light through a diffusive panel, we visualize the free surface below the disk with the total internal reflection technique. The refractive index of Novec 7000 is measured to be 1.25 using the method described in \cite{gluck2011simple} and a minimum liquid height $H =$ 16.5 cm is required to achieve total internal reflection (TIR) for our setup configuration. The reflected free surface is recorded with a high-speed camera at 30k or 48k fps. We synchronize the pressure signal with the high-speed image recording using the procedure outlined in the Supplemental Material \cite{supplemental}, with an accuracy of twice the interframe time of the camera.

In Fig.~\ref{fig:fig1}b, we plot the time evolution of the pressure $P_\text{c}$ measured at the disk center for the smallest ambient temperature reached, $T_0 \approx 11.5^{\circ}$C, for two different impact velocities, namely $U_0 = 0.5$ m/s (red curve) and $U_0 = 2.0$ m/s (blue). Upon impact at $t_\text{a} = 0$, both central pressure signals start to build up prominently, and the time delay is consistent with the impact velocity difference. Subsequently, there is a huge difference in the magnitude, where the maximum pressure reached for $U_0 = 0.5$ m/s, $P_\text{c,max} = 0.084$ bar, is dwarfed by that measured for $U_0 = 2.0$ m/s, namely $P_\text{c,max} = 19.0$ bar, i.e., more than two orders of magnitude larger.

To understand the origin of this huge difference, Fig.~\ref{fig:fig1}c shows a series of representative (re-aspected) snapshots that highlight the distinct behaviour of the entrapped vapor pocket under these different impact conditions. The snapshots are re-aspected to a circular shape, since in the reflected images obtained with the TIR optical set-up that view the free surface at an angle, the disk is deformed into an ellipse. The time after impact $t_\text{a}$ is normalized by the inertial time scale $t_\text{i} = R_0/U_0$ to obtain the non-dimensionalized time $\tilde{t}$. In Fig. \ref{fig:fig1}c, the dark region corresponds to the diffuse reflection of the wetted disk (solid-liquid contact) and the inner bright region within the disk perimeter is the mirror reflection of the liquid surface of the entrapped vapor pocket (solid-vapor contact). From the bottom view images, it is evident that a vapor layer is entrapped under the disk after initial impact at $\tilde{t} = 0.0063$ at both low ($U_0 = 0.5$ m/s) and high ($U_0 = $ 2.0 m/s) impact velocity. The mechanism by which the vapor layer is entrapped is the same as that for air layer entrapment, where local pressure build-up in the {vertically decelerating} intervening gas layer deforms the liquid surface when the solid object is approaching \cite{bouwhuis2015initial,jain2021air}. At low-impact velocity, after impact the liquid-vapor contact line retracts toward the disk center gradually and stably from $\tilde{t} =  0.0063$ to $\tilde{t} =  0.03$ as seen in Fig. \ref{fig:fig1}c(i) and the pocket is punctured at the center only at a very late stage after impact ($t_\text{a} >$ 10 ms as seen in the Supplementary video 1~\cite{supplemental}). In contrast, at high impact velocity, whereas at first the liquid-vapor contact line also starts to retract upon impact, shortly thereafter, (at $\tilde{t} = 0.0125$), the entrapped vapor pocket starts to violently collapse from the center region of the disk, resulting in the dark shaded inner region that expands rapidly with time as shown in Fig. \ref{fig:fig1}c(ii). Estimating the spreading speed $U_{\text{sp}}$ of the vapor pocket from the time evolution of the wetted area, we arrive at $U_{\text{sp}} \approx 600$ m/s, which exceeds the speed of sound in liquid Novec 7000, which lies around $C_\text{sound,L} \approx 500$ m/s \cite{aminian2022ideal}.

\begin{figure}[t!]
\centering

\includegraphics[trim=0cm 0.75cm 0cm 0.5cm,width=0.4\textwidth]{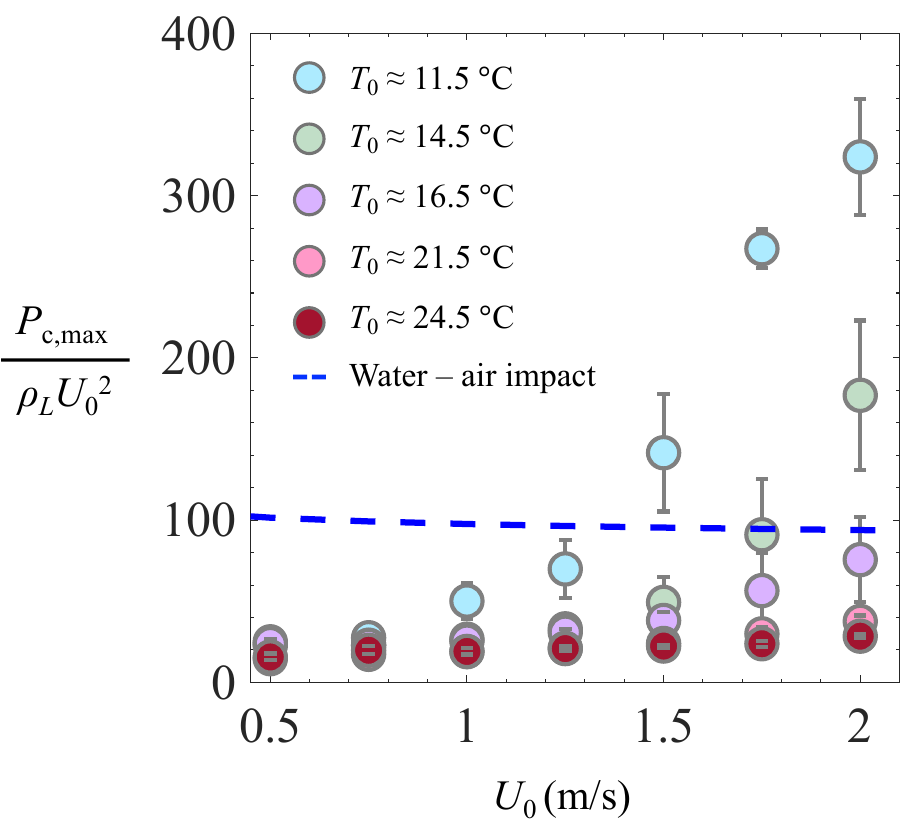}
\caption{\label{fig:fig3} Maximum central impact pressure $P_{\text{c,max}}$, rescaled with the inertial pressure scale $\rho_\text{L} U_0^2$ and plotted against the impact velocity $U_0$ at different ambient temperature $T_0$. Error bars show the standard deviation of the maximum impact pressure at the disk center over at least 5 repetitions of the experiment. The horizontal dashed blue line represents a fit to the data for water--air disk impact from \cite{jain2021air} that follows the classic $\rho_\text{L} U_0^2$ scaling.}
\end{figure}

In general, the maximum local impact pressure $P_{\text{c,max}}$ at the disk center originates from the liquid stagnation pressure and is therefore expected to follow the classic inertial scaling $\rho_\text{L} U_0^2$, which is confirmed experimentally for water-air disk impact \cite{jain2021air}. In Fig.~\ref{fig:fig3} the rescaled maximum local impact pressure at the disk center, $P_{\text{c,max}}/(\rho_\text{L} U_0^2)$, is plotted versus $U_0$ for the boiling liquid case at different values of the ambient temperature $T_0$. At high ambient temperatures $\left( T_0 \approx 21.5 - 24.5 ^{\circ} \text{C} \right)$, also here the data closely follows the $\rho_\text{L} U_0^2$ scaling, evidenced by data lying on a horizontal line in Fig.~\ref{fig:fig3}. Here, the maximum central impact pressures for these higher $T_0$ cases are below the horizontal line due to the larger density difference between Novec 7000 vapour and liquid as compared to that of water and air, resulting in a thicker entrapped pocket under the disk which gives more effective cushioning \cite{fan2025disk}. However, as $T_0$ decreases, we observe a significant pressure increase, causing $P_{\text{c,max}}$ to strongly deviate from the classic $\rho_\text{L} U_0^2$ scaling law. These exceptionally high impact pressures are always recorded when the vapor pocket was observed to collapse rapidly upon impact from the bottom view. Moreover, the deviation is initiated at a lower impact velocity as the ambient temperature decreases. For instance, at $T_0 \approx 16.5^{\circ}\text{C}$, pressures start to deviate at $U_0 =$ 1.5 m/s while $P_{\text{c,max}}$ rises noticeably from $U_0 =$ 1.0 m/s onwards at $T_0 \approx 11.5^{\circ}\text{C}$. 

\begin{figure}
\centering
\includegraphics[width=0.8\columnwidth]{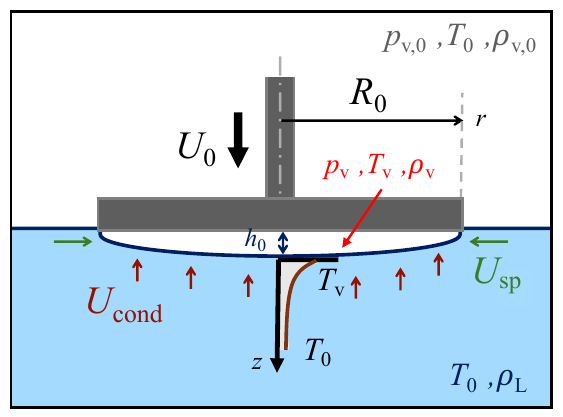}
\caption{\label{fig:fig4} 
Schematic of the model for the collapse of the entrapped vapor pocket upon impact. As the disk descends at constant impact velocity $U_0$ the pressure $p_\text{v}$ and temperature $T_\text{v}$ within the vapor pocket increase, whereas the temperature of the liquid bulk remains at the saturation level ($T_0$). Therefore, to restore equilibrium, the vapor pocket starts to condense at the liquid-vapor interface where the released latent heat is conducted into the liquid. Eventually, this process may cause the vapor pocket to collapse and the wetted area spreads in the radial direction over the disk surface.
}
\end{figure}

We attribute the coinciding strong increase of the impact pressure and the violent collapse of the vapor pocket to a rapid condensation of the vapor inside the pocket, which is a distinguishing characteristic of vapor from air \cite{plesset1977bubble,prosperetti2017vapor}. To rationalize this effect, we assume that upon impact a thin vapor pocket of thickness $h_0$ has been entrapped below the disk, which subsequently becomes pressurized by the liquid, as sketched in Fig.~\ref{fig:fig4}. In response to pressurization, the vapor in the pocket will condense on the liquid-vapor interface, where latent heat is produced that needs to be transported into the liquid \cite{plesset1977bubble}. As further justified in the Supplemental Material \cite{supplemental}, neglecting (i) sensible heat in the vapor, (ii) heat transport through the vapor phase, (iii) density changes in the liquid phase and (iv) condensation on the disk we may write
\begin{equation} \label{eq:balance}
    - L \frac{dm_\text{v}}{dt} = -S_\text{A} k_\text{L} \left. \frac{\partial T}{\partial n} \right|_{\text{interface}} \approx S_\text{A} k_\text{L} \frac{\Delta T_\text{v}}{\delta_{\text{th}}}\,,
\end{equation}
where $L$ is the latent heat of vaporization, $dm_\text{v}/dt$ the rate of change of the vapor mass in the pocket, $S_\text{A}$ the surface area of the pocket, and $k_\text{L} = \alpha_\text{L}\rho_\text{L} C_\text{L}$ the thermal conductivity of the liquid (with $\alpha_\text{L}$ and $C_\text{L}$ the thermal diffusivity and specific heat capacity of the liquid, respectively). The normal temperature gradient $\partial T/\partial n$ at the interface can be estimated as the ratio of the temperature difference $\Delta T_\text{v} = T_\text{v} - T_0$ between vapor and liquid and the   thermal boundary layer thickness $\delta_{\text{th}} \approx \sqrt{\pi \alpha_\text{L} t_\text{c}}$ where $t_\text{c}$ is the collapse time scale of the vapor pocket defined as $t_\text{c} \approx h_0/U_0$. The right hand side of Eq.~\eqref{eq:balance} can be rewritten as
\begin{equation} \label{eq:mass}
    \frac{dm_\text{v}}{dt} \approx - \rho_{\text{v,0}} S_\text{A} U_{\text{cond}}\,,
\end{equation}
where $\rho_{\text{v,0}}$ is the equilibrium vapor density and $U_{\text{cond}}$ is the vapor flux into the liquid interface, i.e., the speed at which vapor condenses on the pocket wall. 

The temperature rise $\Delta T_\text{v}$ in the vapor phase can be estimated from the pressure increase $\Delta p_\text{v}$ upon pressurization using the linearized Clausius-Clapeyron equation
\begin{equation} \label{eq:CC}
   \frac{\Delta T_\text{v}}{T_0} \approx \beta \frac{\Delta p_\text{v}}{p_{\text{v,0}}}\,,
\end{equation}
where $\beta = R_\text{s} T_0/L$, with $R_\text{s}$ the specific gas constant of the vapor, is a small dimensionless parameter. Combining Eqs.~\eqref{eq:balance},~ \eqref{eq:mass}, and~\eqref{eq:CC}, the vapor flux into the liquid interface can be written as
\begin{equation} \label{eq:Ucond/U0}
    \frac{U_{\text{cond}}}{U_0} \sim \frac{\beta^2}{\sqrt{\pi}} \frac{\rho_\text{L}}{\rho_\text{v,0}} \frac{C_\text{L}}{R_\text{s}} \sqrt{\frac{\alpha_\text{L}}{h_0 U_0}}  \frac{\Delta p_\text{v}}{p_\text{v,0}}\,.
\end{equation}

\begin{figure}[t]
\centering
\includegraphics[trim=0cm 0.5cm 0cm .45cm,, width=0.4\textwidth]{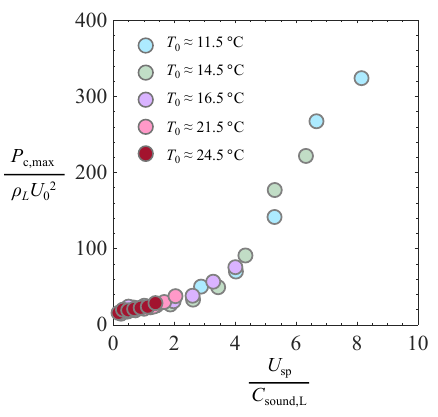}
\caption{\label{fig:fig5} Plotting the rescaled maximum impact pressure $P_{\text{c,max}}/\rho_\text{L}U_0^2$ at the disk center from Fig.~\ref{fig:fig3} against $U_{\text{sp}}/C_{\text{sound,L}}$ as derived from the proposed theoretical model gives a decent collapse of data. The maximum central impact pressure induced during boiling impact can be described as a result of vapor condensation that strongly depends on the impact velocity and vapor density of the boiling system. (Note that an additional data point for $U_0 = 2.25$ m/s at $T_{0} \approx 14.5^{\circ}$C is included here.)}
\end{figure}

The next step is to estimate the pressure rise $\Delta p_\text{v}$ inside the vapor pocket, by examining in detail what happens just after impact: During the first few microseconds, the liquid just squeezes the vapor inside the pocket with pressure still insufficiently large to provoke a response in the liquid. As soon as the pressure reaches the water hammer value, $\rho_\text{L}C_\text{sound,L}U_0$, the liquid will start to be accelerated by the pressurized vapor and an added mass region will build up below the disk. This so-called water hammer stage \cite{mayer2018flat,korobkin2006numerical} lasts until it reaches the size of the disk, i.e., $t_a \approx R_0/C_{\text{sound,L}} \approx 80$ $\mu$s. Afterwards, the liquid added mass continues to squeeze the vapor pocket potentially leading to further pressure rise. Evaluating the vapor pocket collapse time as in the order of $100$ $\mu$s (refer to Fig. \ref{fig:fig1}c(ii)), locates it in the early, water-hammer stage of impact and consequently the appropriate pressure scale to use in Eq.~\eqref{eq:Ucond/U0} is $\Delta p_\text{v} \approx \rho_\text{L}C_\text{sound,L}U_0$. This is further motivated in the Supplementary Material \cite{supplemental}. Note that, the central impact pressure recorded when the vapor pocket collapses can even exceed the water hammer pressure.

Finally, while condensation occurs along the entire vapor pocket interface, the collapse of the entrapped vapor pocket is determined by how fast the liquid-solid contact area spreads in the plane of the disk surface. Using continuity to relate the spreading speed of the liquid-solid contact area $U_\text{sp}$ over the disk surface to
$U_{\text{cond}}$, namely $2 U_{\text{sp}} h_0 \sim U_{\text{cond}} R_0$, we obtain
\begin{equation} \label{eq:Uin/Csound}
    \frac{U_{\text{sp}}}{C_{\text{sound,L}}} \sim \frac{\beta^2}{2\sqrt{\pi}} \left( \frac{R_0}{h_0} \right)^{\!\!3/2} \!\!\!\frac{\rho_\text{L}}{\rho_\text{v,0}} \frac{C_\text{L}}{R_\text{s}} \sqrt{\frac{\alpha_\text{L}}{R_0 U_0}}  \frac{\rho_\text{L}U_0^2}{p_\text{v,0}}\,.
\end{equation}
In this equation, the only remaining unknown quantity is the initial thickness $h_0$ of the vapor pocket. Prior to impact, when the disk is moving towards the free liquid surface, pressure builds up as $\sim \rho_\text{v} U_0^2$ underneath the impacting disk as the intervening vapor layer between the disk and the liquid surface is being squeezed. This causes the liquid surface to deform, with the surface under the disk center being depressed the most \cite{mayer2018flat,jain2021KH} and the surface near the disk edge being lifted due to Kelvin-Helmholtz instability \cite{jain2021KH}. Therefore, as illustrated in Fig. \ref{fig:fig4}, upon impact, the disk will first contact with the deformed liquid surface near the disk edge and entrap a thin layer of vapor with an initial radius $R_0$ and central thickness $h_0$ of the order of hundreds of microns. In the potential flow framework \cite{peters2013splash,bouwhuis2015initial}, one expects the deformation of the liquid interface to be proportional to the vapor to liquid density ratio, such that $h_0/R_0 = \lambda\rho_\text{v,0}/\rho_\text{L}$, where $\lambda$ is a numerical constant of order unity. Inserting this expression in Eq.~\eqref{eq:Uin/Csound} gives 
\begin{equation} \label{eq:Uin/Csound final}
    \frac{U_{\text{sp}}}{C_{\text{sound,L}}} \sim \frac{\beta^2}{2\sqrt{\pi}\lambda^{3/2}} \left(\frac{\rho_{\text{L}}}{\rho_{\text{v,0}}} \right)^{\!\!5/2}\!\! \frac{C_\text{L}}{R_\text{s}} \sqrt{\frac{\alpha_\text{L}}{R_0 U_0}}  \frac{\rho_\text{L}U_0^2}{p_\text{v,0}}\,.
\end{equation}
In the above expression the dependence on the impact velocity $U_0$ is clear since $U_{\text{sp}} \sim U_0^{3/2}$. Lowering the ambient temperature $T_0$, the dominant effect is in the reduced vapor density $\rho_{\text{v,0}}$, which influences $U_{\text{sp}}$ both directly and through a smaller initial thickness $h_0$ of the entrapped vapor layer. 

We then compute $U_{\text{sp}}/C_{\text{sound,L}}$ using values of the temperature-dependent properties in Eq. \eqref{eq:Uin/Csound final} from \cite{aminian2022ideal,widiatmo2001equations,rausch2015density}, see Supplementary Material \cite{supplemental}. The fitting parameter $\lambda$ is taken to be $2.1$ as obtained by fitting the experimentally measured $h_0$ (not shown here, but see \cite{fan2025disk}). Replotting the rescaled maximum impact pressure at the disk center from Fig. \ref{fig:fig3} against the computed $U_{\text{sp}}/C_{\text{sound,L}}$ gives a decent collapse of data as shown in Fig. \ref{fig:fig5}, suggesting that this simple model predicts the dependence of the vapor condensation in the pocket on the impact velocity and ambient temperature during boiling liquid impact. In addition, this confirms that the condensation-induced collapse of the vapor pocket is responsible for the very high impact pressures exerted on the impacting surface.

The mechanism by which condensation creates the anomalously high maximum central impact pressures is the following: When phase change does not play a dominant role, the solid-liquid impact is cushioned by the (relatively slow) pressurization of the vapor pocket. If the vapor condenses sufficiently fast, the liquid reaches the solid interface unbraked, resulting in a large and sudden impact pressure peak. 

In summary, we experimentally observed anomalously high impact pressures during impact of a disk onto a boiling liquid, exceeding expected values by more than an order of magnitude. These high pressures are observed to coincide with a rapid collapse of the vapor pocket entrapped below the disk, and found to occur for low equilibrium vapor densities and high impact velocities. We constructed a scaling model accounting for our findings that provides quantitative insight in the material parameter values for which anomalous behaviour may occur. Most notably, our experiments show that phase change occurring during impact under thermal equilibrium conditions may lead to qualitatively different physics compared to that encountered during impact experiments in non-condensable gases such as air. Clearly, vapor condensation during boiling liquid impact is detrimental to the effectiveness of gas cushioning and may lead to unexpectedly high local impact pressure on the solid surface upon impact. Therefore, it is imperative to account for phase change during boiling liquid impact while designing a containment system for cryogenic fuels to ensure the safety and integrity of the tank. \\

We thank G.-W. Bruggert, D. van Gils, M. Bos and T. Zijlstra for their technical support. We would also like to thank U. Jain for providing his original experimental data. This work is part of the Vici project IMBOL (project number 17070) which is partly financed by the Dutch Research Council (NWO). \\

\noindent{\textit{Data availability:} The data that support the findings of this article are openly available \cite{dataSet}.}

\bibliographystyle{apsrev4-2}
\bibliography{IMBOLreflist}

@PREAMBLE{
 "\providecommand{\noopsort}[1]{}" 
 # "\providecommand{\singleletter}[1]{#1}%" 
}

@string{JFM = "J. Fluid Mech."}

@string{Fluidphaseequilibria = "Fluid Ph. Equilib."}

@string{JournalofChemicalEngineeringData = "J. Chem. \& Eng. Data."}

@string{Journaloffluidsandstructures = "J. Fluids Struct."}

@string{PhysicalReviewFluids  = "Phys. Rev. Fluids"}

@string{PhysicalReviewLetters  = "Phys. Rev. Lett."}

@string{ARFM = "Annu. Rev. Fluid Mech."}

@article{widiatmo2001equations,
  title={Equations of state for fluorinated ether refrigerants, pentafluoroethyl methyl ether and heptafluoropropyl methyl ether},
  author={Widiatmo, Januarius V and Watanabe, Koichi},
  journal=Fluidphaseequilibria,
  volume={183},
  pages={31--39},
  year={2001},
  publisher={Elsevier}
}

@article{ohta2001liquid,
  title={Liquid-phase thermodynamic properties of new refrigerants: Pentafluoroethyl methyl ether and heptafluoropropyl methyl ether},
  author={Ohta, Hirofumi and Morimoto, Yoshiyuki and Widiatmo, Januarius V and Watanabe, Koichi},
  journal=JournalofChemicalEngineeringData,
  volume={46},
  number={5},
  pages={1020--1024},
  year={2001},
  publisher={ACS Publications}
}

@article{mayer2018flat,
  title={Flat plate impact on water},
  author={Mayer, Hans C and Krechetnikov, Rouslan},
  journal=JFM,
  volume={850},
  pages={1066--1116},
  year={2018},
  publisher={Cambridge University Press}
}

@article{jain2021air,
  title={Air entrapment and its effect on pressure impulses in the slamming of a flat disc on water},
  author={Jain, Utkarsh and Vega-Mart{\'\i}nez, Patricia and Van Der Meer, Devaraj},
  journal=JFM,
  volume={928},
  pages={A31},
  year={2021},
  publisher={Cambridge University Press}
}

@article{hicks2018lng,
  title={LNG-solid impacts with gas cushioning and phase change},
  author={Hicks, Peter D},
  journal=Journaloffluidsandstructures,
  volume={80},
  pages={22--36},
  year={2018},
  publisher={Elsevier}
}

@article{lee2021experimental,
  title={Experimental study on effect of density ratio and phase transition during sloshing impact in rectangular tank},
  author={Lee, Jeoungkyu and Ahn, Yangjun and Kim, Yonghwan},
  journal={Ocean Eng.},
  volume={242},
  pages={110105},
  year={2021},
  publisher={Elsevier}
}

@article{bagnold1939interim,
  title={INTERIM REPORT ON WAVE-PRESSURE RESEARCH.},
  author={Bagnold, RA},
  journal={J. Civ. Eng.},
  volume={12},
  number={7},
  pages={202--226},
  year={1939},
  publisher={Thomas Telford-ICE Virtual Library}
}

@article{peters2013splash,
  title={Splash wave and crown breakup after disc impact on a liquid surface},
  author={Peters, Ivo R and van der Meer, Devaraj and Gordillo, JM},
  journal=JFM,
  volume={724},
  pages={553--580},
  year={2013},
  publisher={Cambridge University Press}
}

@article{rausch2015density,
  title={Density, surface tension, and kinematic viscosity of hydrofluoroethers HFE-7000, HFE-7100, HFE-7200, HFE-7300, and HFE-7500},
  author={Rausch, Michael H and Kretschmer, Lorenz and Will, Stefan and Leipertz, Alfred and Froba, Andreas P},
  journal=JournalofChemicalEngineeringData,
  volume={60},
  number={12},
  pages={3759--3765},
  year={2015},
  publisher={ACS Publications}
}

@article{fan2024air,
  title={Air-cushioning below an impacting wave-structured disk: Free-surface deformation and slamming load},
  author={Fan, Yee Li and Jain, Utkarsh and Van Der Meer, Devaraj},
  journal=PhysicalReviewFluids,
  volume={9},
  number={1},
  pages={010501},
  year={2024},
  publisher={APS}
}

@article{fan2025disk,
  title={Disk impact on a boiling liquid: Dynamics of the entrapped vapor pocket},
  author={Fan, Yee Li and Palacios Mu{\~n}iz, Bernardo and Kim, Nayoung and van der Meer, Devaraj},
  journal= PhysicalReviewFluids,
  volume={10},
  number={10},
  pages={100505},
  year={2025},
  publisher={APS}
}

@article{verhagen1967impact,
  title={The impact of a flat plate on a water surface},
  author={Verhagen, JHG},
  journal={J. Sh. Res},
  volume={11},
  number={04},
  pages={211--223},
  year={1967},
  publisher={SNAME}
}

@article{maillard2009influence,
  title={Influence of DR between liquid and gas on sloshing model test results},
  author={Maillard, S and Brosset, L},
  journal={IJOPE},
  volume={19},
  number={04},
  year={2009},
  publisher={OnePetro}
}

@article{hicks2012air,
  title={Air trapping at impact of a rigid sphere onto a liquid},
  author={Hicks, PD and Ermanyuk, EV and Gavrilov, NV and Purvis, R},
  journal=JFM,
  volume={695},
  pages={310--320},
  year={2012},
  publisher={Cambridge University Press}
}

@article{tran2013air,
  title={Air entrainment during impact of droplets on liquid surfaces},
  author={Tran, Tuan and de Maleprade, H{\'e}l{\`e}ne and Sun, Chao and Lohse, Detlef},
  journal=JFM,
  volume={726},
  pages={R3},
  year={2013},
  publisher={Cambridge University Press}
}

@article{lee2012does,
  title={How does an air film evolve into a bubble during drop impact?},
  author={Lee, Ji San and Weon, Byung Mook and Je, Jung Ho and Fezzaa, Kamel},
  journal=PhysicalReviewLetters,
  volume={109},
  number={20},
  pages={204501},
  year={2012},
  publisher={APS}
}

@article{hendrix2016universal,
  title={Universal mechanism for air entrainment during liquid impact},
  author={Hendrix, Maurice HW and Bouwhuis, Wilco and van der Meer, Devaraj and Lohse, Detlef and Snoeijer, Jacco H},
  journal=JFM,
  volume={789},
  pages={708--725},
  year={2016},
  publisher={Cambridge University Press}
}

@article{bouwhuis2015initial,
  title={Initial surface deformations during impact on a liquid pool},
  author={Bouwhuis, Wilco and Hendrix, Maurice HW and van der Meer, Devaraj and Snoeijer, Jacco H},
  journal=JFM,
  volume={771},
  pages={503--519},
  year={2015},
  publisher={Cambridge University Press}
}

@article{bouwhuis2012maximal,
  title={Maximal air bubble entrainment at liquid-drop impact},
  author={Bouwhuis, Wilco and van der Veen, Roeland CA and Tran, Tuan and Keij, Diederik L and Winkels, Koen G and Peters, Ivo R and van der Meer, Devaraj and Sun, Chao and Snoeijer, Jacco H and Lohse, Detlef},
  journal=PhysicalReviewLetters,
  volume={109},
  number={26},
  pages={264501},
  year={2012},
  publisher={APS}
}

@article{carrat2023air,
  title={Air entrapment at impact of a conus onto a liquid},
  author={Carrat, J-B and Gavrilov, N and Cherdantsev, A and Shmakova, N and Ermanyuk, E},
  journal=JFM,
  volume={966},
  pages={R1},
  year={2023},
  publisher={Cambridge University Press}
}

@article{marston2011bubble,
  title={Bubble entrapment during sphere impact onto quiescent liquid surfaces},
  author={Marston, JO and Vakarelski, Ivan Uriev and Thoroddsen, Sigurdur T},
  journal=JFM,
  volume={680},
  pages={660--670},
  year={2011},
  publisher={Cambridge University Press}
}

@article{kim2021water,
  title={Water impact of a surface-patterned disk},
  author={Kim, Taehyun and Kim, Donghyun and Kim, Daegyoum},
  journal=JFM,
  volume={915},
  pages={A52},
  year={2021},
  publisher={Cambridge University Press}
}

@article{ermanyuk2011experimental,
  title={Experimental study of disk impact onto shallow water},
  author={Ermanyuk, EV and Gavrilov, NV},
  journal={J. Appl. Mech. Tech. Phys.},
  volume={52},
  pages={889--895},
  year={2011},
  publisher={Springer}
}

@article{josserand2016drop,
  title={Drop impact on a solid surface},
  author={Josserand, Christophe and Thoroddsen, Sigurdur T},
  journal=ARFM,
  volume={48},
  number={1},
  pages={365--391},
  year={2016},
  publisher={Annual Reviews}
}

@misc{3MNovec,
  howpublished = {3M$^{\text{TM}}$ Novec$^{\text{TM}}$ 7000 Engineered Fluid, website \textsf{\scriptsize https://www.3m.com/3M/en\_US/p/d/b5005006004/}
  }
}

@article{perkins2022measurement,
  title={Measurement and Correlation of the Thermal Conductivity of 1, 1, 1, 2, 2, 3, 3-Heptafluoro-3-methoxypropane (RE-347mcc)},
  author={Perkins, Richard A and Huber, Marcia L and Assael, Marc J},
  journal={Int. J. Thermophys.},
  volume={43},
  pages={1--16},
  year={2022},
  publisher={Springer}
}

@article{jain2021KH,
  title={Air-cushioning effect and Kelvin-Helmholtz instability before the slamming of a disk on water},
  author={Jain, Utkarsh and Gauthier, Ana{\"\i}s and Lohse, Detlef and van Der Meer, Devaraj},
  journal=PhysicalReviewFluids,
  volume={6},
  number={4},
  pages={L042001},
  year={2021},
  publisher={APS}
}

@article{korobkin2006numerical,
  title={Numerical study of jet flow generated by impact on weakly compressible liquid},
  author={Korobkin, Alexander A and Iafrati, Alessandro},
  journal={Phys. Fluids},
  volume={18},
  number={3},
  year={2006},
  publisher={AIP Publishing}
}

@article{aminian2022ideal,
  title={Ideal Gas Heat Capacity and Critical Properties of HFE-Type Engineering Fluids: Ab Initio Predictions of C p ig, Modeling of Phase Behavior and Thermodynamic Properties Using Peng--Robinson and Volume-Translated Peng--Robinson Equations of State},
  author={Aminian, Ali and Celn{\`y}, David and Mickoleit, Erik and J{\"a}ger, Andreas and Vin{\v{s}}, V{\'a}clav},
  journal={Int. J. Thermophys.},
  volume={43},
  number={6},
  pages={87},
  year={2022},
  publisher={Springer}
}

@article{gluck2011simple,
  title={A simple method to measure the refractive index of a liquid},
  author={Gluck, Paul},
  journal={Phys. Educ.},
  volume={46},
  number={3},
  pages={253},
  year={2011},
  publisher={IOP Publishing}
}

@article{prosperetti2017vapor,
  title={Vapor bubbles},
  author={Prosperetti, Andrea},
  journal=ARFM,
  volume={49},
  number={1},
  pages={221--248},
  year={2017},
  publisher={Annual Reviews}
}

@article{dias2018slamming,
  title={Slamming: Recent progress in the evaluation of impact pressures},
  author={Dias, Fr{\'e}d{\'e}ric and Ghidaglia, Jean-Michel},
  journal=ARFM,
  volume={50},
  number={1},
  pages={243--273},
  year={2018},
  publisher={Annual Reviews}
}

@article{plesset1977bubble,
  title={Bubble dynamics and cavitation},
  author={Plesset, Milton S and Prosperetti, Andrea},
  journal=ARFM,
  volume={9},
  pages={145--185},
  year={1977}
}

@misc{dataSet,
  howpublished = {Y. L. E. Fan, ‘Dynamic pressure enhancement upon disk impact on a boiling liquid’ [Data set], Zenodo (2025), https://doi.org/10.5281/zenodo.17233929}
  }

@misc{supplemental,
  howpublished = {See Supplemental Material at [URL will be inserted by publisher] for the procedure we adopted to synchronize the pressure signal with the high-speed image recording, the temperature-dependent properties of Novec 7000 used for computation of Eq. \eqref{eq:Uin/Csound final}, the justification of the heat-mass balance equation (Eq. \eqref{eq:balance}) used in our model, the justification on using the water hammer pressure as the relevant pressure scale in Eq. \eqref{eq:Ucond/U0}, a comparison of inertial and compressive rescaling of the impact pressure, and the movies of the original experimental high-speed recordings of the impact process viewed from the bottom showed in Fig. \ref{fig:fig1}c, which includes Refs. \cite{Ancellin2012,Brosset2013,brennenbook,muniz2024impact}}
  }

@InProceedings{Ancellin2012,
  author       = "Ancellin, M. and Ghidaglia, J. M. and Brosset, L.",
  title            = "{Influence of phase transition on sloshing impact pressures described by a generalized Bagnold's model}",
  booktitle    = "Proc. of the 22nd ISOPE",
  series        = "Rhodes, Greece, June 17-22, 2012",
  year           = "2012",
  pages        =  "{ISOPE-I-12-372}",
  volume      = {3}
  }

@InProceedings{Brosset2013,
  author       = "Brosset, L. and Ghidaglia, J. M. and Le Tarnec, L. and Guilcher, P. M.",
  title            = "{Generalized Bagnold model}",
  booktitle    = "Proc. of the 23rd ISOPE",
  series        = "Anchorage, Alaska USA, June 30-July 5, 2013",
  year           = "2013",
  pages        = "{ISOPE-I-13-331}",
  volume      = {3}
  }

@misc{brennenbook,
  title           = {An Internet Book on Fluid Dynamics},
  author      = {Brennen, C.A.},
  year         = 2006,
  note         = {\url{http://brennen.caltech.edu/fluidbook} [Accessed: March 25,2025]}
}

@phdthesis{muniz2024impact,

author = "{Palacios Mu{\~n}iz}, Bernardo",
year = "2024",
month = oct,
day = "16",
doi = "10.3990/1.9789036563192",
isbn = "978-90-365-6318-5",
publisher = "University of Twente",
address = "Netherlands",
school = "University of Twente",
}

 \end{document}


\title{Supplemental Material\\{\small Dynamic pressure enhancement upon disk impact on a boiling liquid}\\}

\author{Yee Li (Ellis) Fan$^1$}
\email{Contact author: ellisfan179@gmail.com}
\author{Bernardo Palacios Muñiz$^1$}
\author{Nayoung Kim$^{1}$} 
\author{Devaraj van der Meer$^1$}
\email{Contact author: d.vandermeer@utwente.nl}

\affiliation{%
$^1$Physics of Fluids Group and Max Planck Center Twente for Complex Fluid Dynamics,
MESA+ Institute and J. M. Burgers Centre for Fluid Dynamics, University of Twente,
P.O. Box 217, 7500AE Enschede, The Netherlands}%

\date{\today}

\maketitle

\section{Synchronization of the pressure signal with the high-speed image recording}

\begin{figure}[h]
    \centering
    \includegraphics[width=.8\linewidth]{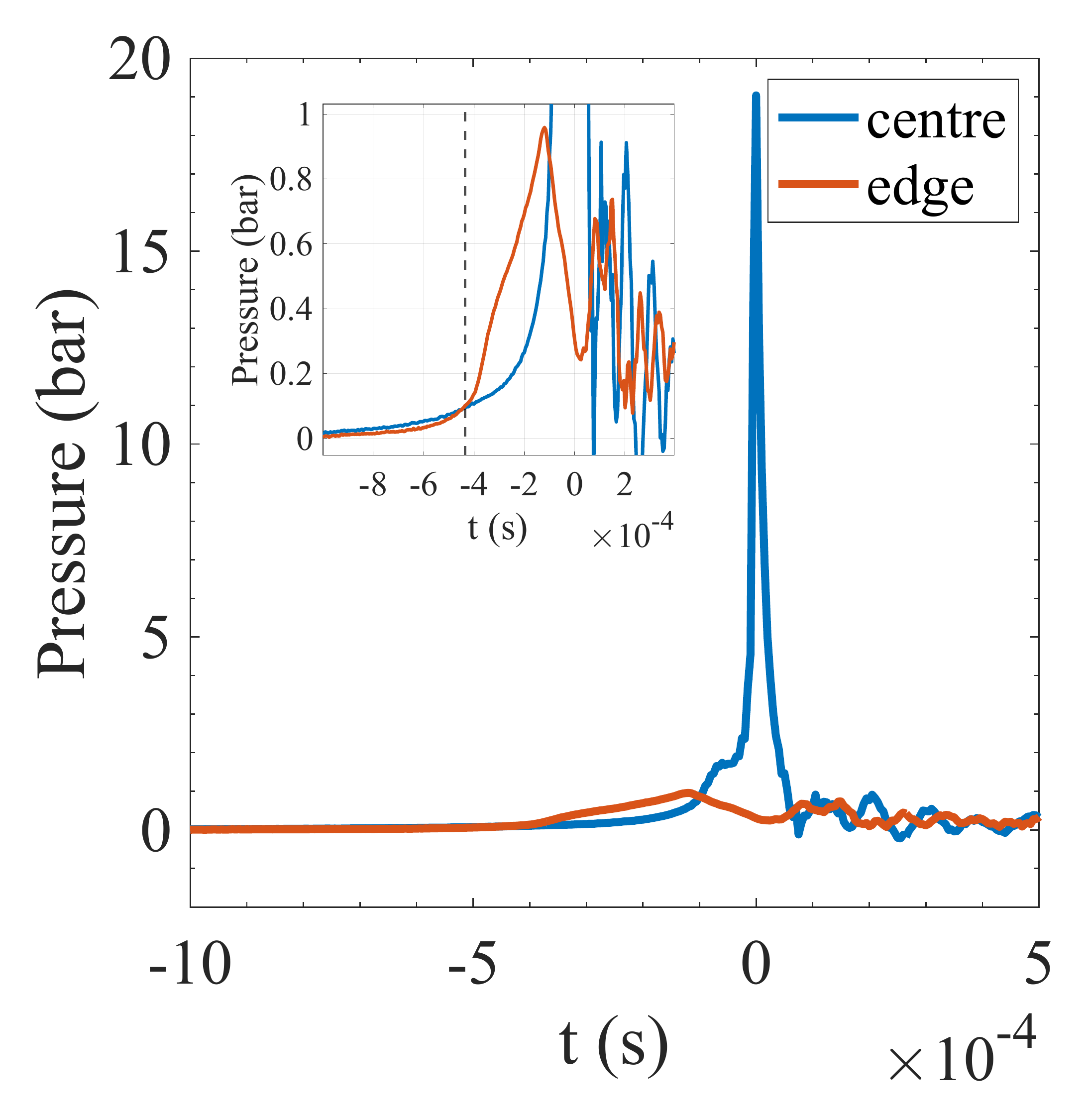}
    \caption{Pressure signals recorded by the pressure sensors at the disc centre (blue) and near the disc edge (orange) for the case in Fig. 1c(ii) in the main text, centered at $t = 0$ when the maximum pressure is reached at the disc center. The inset shows the zoomed-in view of the edge pressure signal. The vertical black dashed line indicates the time at which the signal at the reference pressure sensor near the disk edge rises above $10\%$ of its maximum value, which coincides with the first contact between the liquid and the rim of the disc observed in the high-speed images.}
    \label{fig:sync}
\end{figure}

In our experiments, we did not directly synchronize the pressure signal with the high-speed image recording signal due to the different ways in which the signals are triggered. To nevertheless synchronize the two signals in time, we used the following procedure. Fig. \ref{fig:sync} shows the typical pressure signals received by the pressure sensors at the disc center and near the disc edge, at high impact velocity and low ambient temperature, where the vapor pocket collapses. In these experiments, we observe that when we align the pressure maximum in the center sensor on the disc with the complete collapse in the images (i.e., 0.42 ms in Fig. 1c(ii)), in all cases, the signal received by a reference pressure sensor near the disk edge is found to start increasing significantly when the first contact between the liquid and solid takes place near the rim of the disc. Physically, this stands to reason, since one would expect the pressure in the vapor pocket to start rising as soon as the pocket closes at the disk edge.

Therefore, we decided to define $t_a=0$ as the moment of first contact in the images and the time at which the signal at the reference pressure sensor near the disk edge rises above $10\%$ of its maximum value (e.g., the vertical black dashed line in the inset of Fig. \ref{fig:sync}). Using this definition, for the cases where there is a high pressure and a vapor pocket collapse, the maximum in the pressure and the vapor pocket collapse coincide to within 40 -- 70 $\mu$s, corresponding to twice the interframe time of the camera (frame rate ranging from 30k to 48k fps), which we therefore define as the accuracy of our synchronization.

\section{Temperature-dependent properties of Novec 7000}

\begin{figure}[h]
    \centering
    \includegraphics[width=1\linewidth]{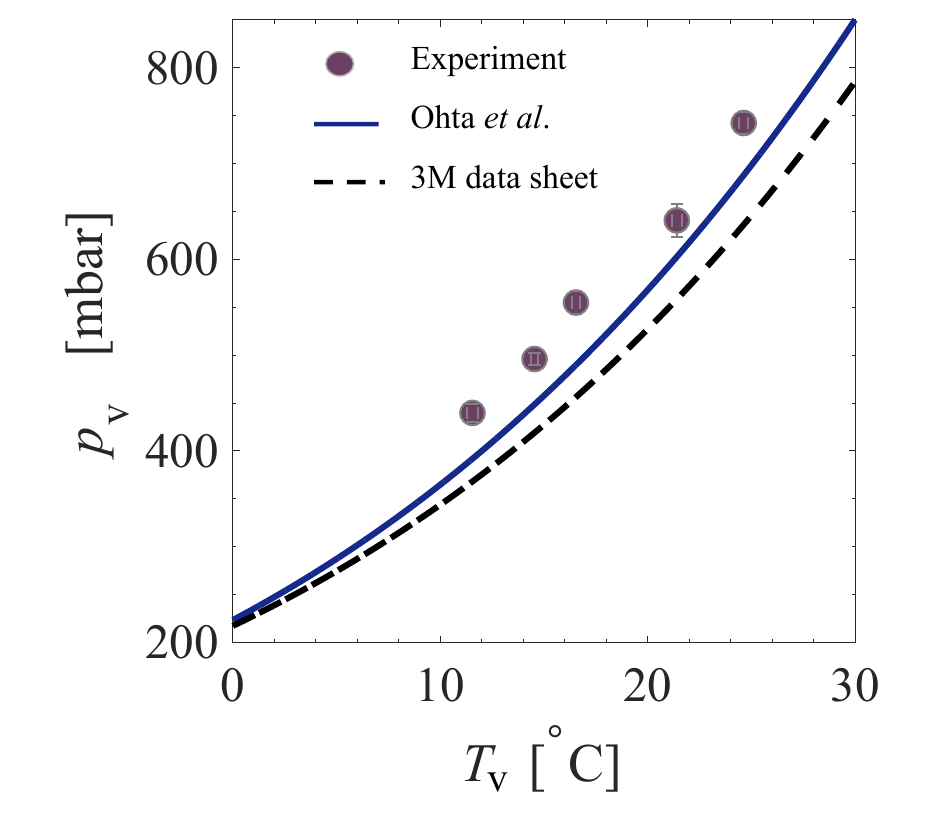}
    \caption{Vapor curve of Novec 7000, representing the saturation vapor pressure $p_\text{v}$ as a function of saturation temperature $T_\text{v}$, computed using the correlation Eq. \ref{eq:supp pv0} from \cite{ohta2001liquid}. Circular points indicate the equilibrium temperatures $T_{0}$ in our experiment and their corresponding averaged vapor pressures $p_\text{v,0}$. The dashed black line is the vapor curve provided in the datasheet from the manufacturer 3M \cite{3MNovec}.}
    \label{fig:vapCurve_supp}
\end{figure}

Here, we clarify the way we obtain the temperature-dependent physical properties of Novec 7000 from several references \cite{3MNovec,widiatmo2001equations,ohta2001liquid,rausch2015density,perkins2022measurement,aminian2022ideal}. Wherever available, we have used experimentally well-validated sources for these Novec 7000 \cite{widiatmo2001equations,ohta2001liquid,rausch2015density,perkins2022measurement}, but in the absence of such we have used theoretically predicted data \cite{aminian2022ideal}.

\begin{figure*}
\includegraphics[width=.8\textwidth]{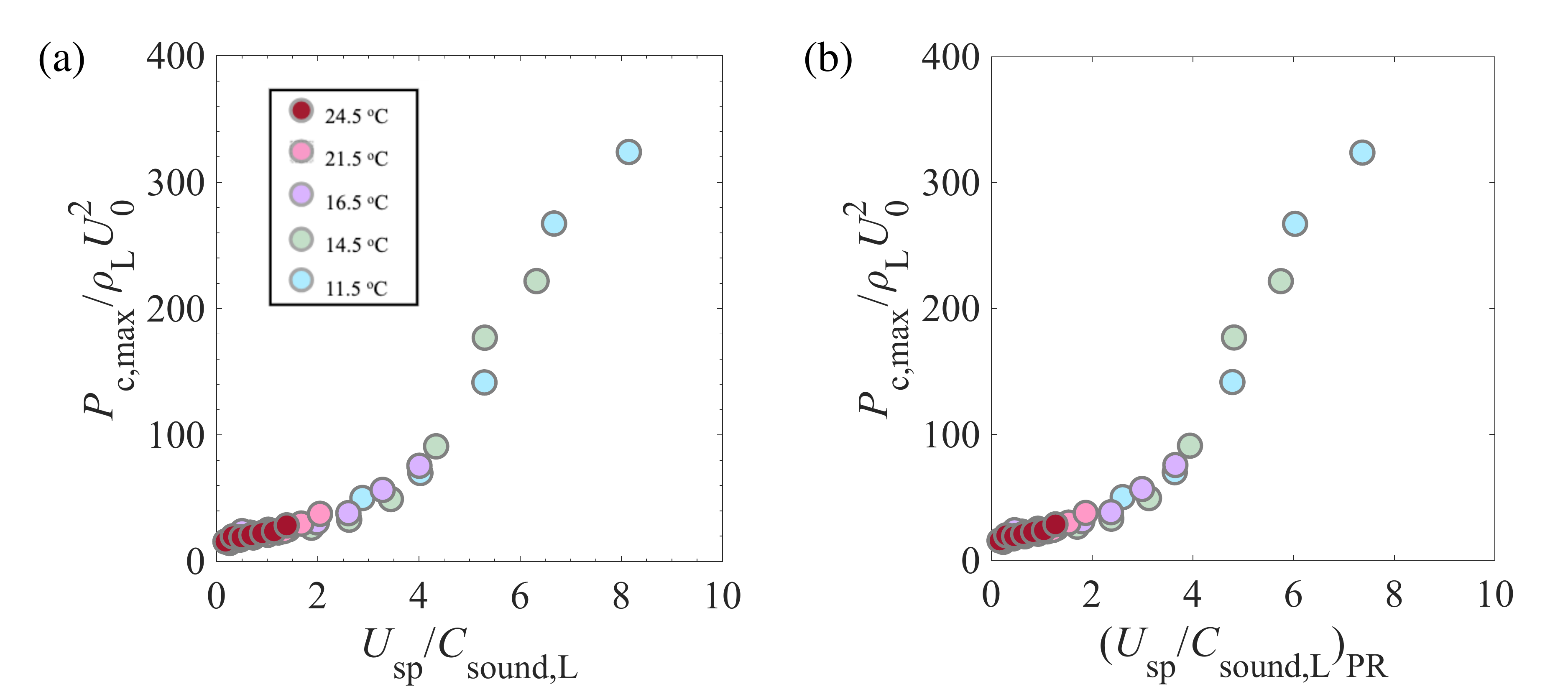}
\caption{\label{fig:Sup_plots}  Comparison between the plot of $P_\text{c,max}/\rho_\text{L}U_0^2$ against $U_\text{sp}/C_\text{sound,L}$ computed with $\rho_\text{v,0}$ based on (a) ideal-gas law as in Fig. 4 in the main text and (b) Peng-Robinson Equation of State.}
\end{figure*}

The saturation pressure $p_\text{v}$ for Novec 7000 at different (absolute) saturation temperatures $T_\text{v}$ is computed based on the empirical and experimentally confirmed correlation provided in \cite{ohta2001liquid}:
\begin{equation} \label{eq:supp pv0}
    \text{ln} \ P_\text{r} = \frac{T_\text{c}}{T_\text{v}} \left( a_1\tau +a_2\tau^{1.5} +a_3\tau^3 + a_4\tau^6 \right)
\end{equation}
where $P_\text{r} = P_\text{v}/P_\text{c}$ and $\tau = 1 - T_\text{v}/T_\text{c}$. Here, subscript c indicates the critical temperature $T_\text{c}$ and critical pressure $P_\text{c}$. For Novec 7000, the values are $P_\text{c} = 2476$ kPa, $T_\text{c} = 437.7$ K, $a_1 = -7.951$ , $a_2 = 1.510$, $a_3 = -4.481$ and $a_4 = -20.835$. The vapor curve of Novec 7000 obtained based on this correlation (with $T_\text{v}$ converted to the Celcius scale) is plotted as a solid line in Fig. A1, together with the averaged equilibrium pressure $p_\text{v,0}$ obtained in our experiments (circles) for the corresponding equilibrium temperature $T_{0}$. This correlation is used as the reference, rather than the one provided in the 3M data sheet \cite{3MNovec} (dashed line in Fig. A1) because repetitive measurements of the saturation pressure of Novec 7000 at various temperatures in \cite{muniz2024impact} (with a lower volume of Novec 7000 than was used in our experiment) consistently coincide with the one provided in \cite{ohta2001liquid}. With this, the equilibrium pressure $p_\text{v,0} \equiv p_\text{v}(T_{0})$ is computed with Eq. \ref{eq:supp pv0}. Using the ideal gas law, we calculate the equilibrium vapor density $\rho_\text{v,0}$ as
\begin{equation}
   \rho_\text{v,0} = \frac{p_\text{v,0}}{R_\text{s}T_{0}}
\end{equation}
where the specific gas constant $R_\text{s} = R_\text{u}/M$, with $R_\text{u} = 8.314$ J/mol K the universal gas constant, $M$ = 0.2 kg/mol the molar mass of Novec 7000, and $T_0$ expressed in the Kelvin scale. Note that the behaviour of the vapor may become non-ideal at the phase boundary and deviate from the ideal gas law, which may result in a larger equilibrium vapor density than what is predicted by the ideal gas law. Comparing the equilibrium vapor density predicted by the ideal gas law (see Table \ref{table:NovecProp}) with that predicted by using the Peng–Robinson equation of state (EOS) \cite{aminian2022ideal}, the equilibrium vapor density is around 8-10$\%$ larger. While this slightly lowers the value of $U_\text{sp}/C_\text{sound,L}$ in Fig.~4 in the main text, the collapse of the data is not affected, as shown in the plots in Fig. \ref{fig:Sup_plots}. There is a caveat in applying the Peng–Robinson EOS, namely that the vapor curve used in that formalism \cite{aminian2022ideal} is less accurate than Eq.~\eqref{eq:supp pv0}. If one naively uses the latter vapor curve to compute vapor densities with the Peng–Robinson EOS, the result is much closer to the vapor density derived from the ideal gas law. %

The properties of liquid Novec 7000 are mainly derived from \cite{aminian2022ideal}, which comes with a table containing data of the thermodynamic properties along the saturation lines for Novec 7000 obtained using the Peng-Robinson equation of state, which is a cubic equation of state. We interpolate the provided data with cubic interpolation to obtain the latent heat of vaporization $L$, the specific heat capacity $C_\text{L}$ at constant pressure, the density $\rho_\text{L}$ and the speed of sound $C_\text{sound,L}$ in the liquid phase, corresponding to our experimental equilibrium temperature $T_0$. Meanwhile, the thermal conductivity of liquid Novec 7000 $k_\text{L}$ is calculated with the formula provided in the 3M data sheet \cite{3MNovec}
\begin{equation}
    k_\text{L} = 0.0789 - 0.000196T_0
\end{equation}
which coincides well with the measurements provided in \cite{perkins2022measurement}. Note that here $T_0$ is in $^{\circ}$C. The thermal diffusivity of liquid Novec 7000 $\alpha_\text{L}$ can then be calculated as:
\begin{equation}\label{eq:alpha}
    \alpha_\text{L} = \frac{k_\text{L}}{\rho_\text{L}C_\text{L}}
\end{equation}

\begin{sidewaystable}[h!]
\centering
\begin{tabular}{|c| c| c| c| c| c| c| c| c| c| c| c| c| c| c| c| c|} 
\hline
 $T_{0}$ $[^{\circ}\text{C}]$ & 
 $p_\text{v,0} [\text{Pa}]$ & 
 $\rho_{\text{v,0}} [\text{kg/m}^3]$ & 
 $\rho_{\text{L}} [\text{kg/m}^3]$ & 
 $C_\text{L}$ [J/kg K] & 
 $k_{\text{L}}$ [W/m K] & 
 $L$ [kJ/kg] & 
 $C_{\text{sound,L}}$ [m/s] & 
 $\alpha_\text{L}$ [m$^2$/s] & 
 $\mu_\text{v}$ [Pa/s]  & 
 $\nu_{\text{L}}$ [m$^2$/s] &
 $c_\text{p,v}$ [J/kg K] & 
 $k_\text{v}$ [W/m K] & 
 $\alpha_\text{v}$ [m$^2$/s] & 
 $\gamma$
  \\ [0.5ex] 

 \hline\hline
11.5 &	39105.9 &	3.3046 & 1450.17 & 1140.47 &	0.07755& 141.5 &	497.37& 0.4689e-07 &	1.1234e-05 & 3.5866e-07 & 863.25 & 0.0109 &	0.3823e-05 & 1.0554 \\ 
14.5 &	44784.3 &	3.7450 & 1442.67 & 1147.68 &	0.07696& 140.5 &	487.11& 0.4648e-07 &	1.1362e-05 & 3.4790e-07 & 869.17 & 0.0111 &	0.3418e-05 & 1.0555 \\
16.5 &	48913.6 &	4.0620 & 1437.60 & 1152.55 &	0.07656& 139.9&	480.36& 0.4621e-07 &	1.1448e-05 & 3.4147e-07 & 873.12 & 0.0113 &	0.3178e-05 & 1.0556 \\
21.5 &	60537.9 &	4.9421 & 1424.63 & 1164.83 &	0.07559& 138.3 &	463.74& 0.4555e-07 &	1.1665e-05 & 3.2378e-07	& 883.09 & 0.0116 &	0.2667e-05 & 1.0561 \\
24.5 &	68474.3	&   5.5337 & 1416.60 & 1172.34 &	0.07500& 137.3 &	453.95& 0.4516e-07 &	1.1797e-05 & 3.1142e-07 & 889.11 & 0.0119 &	0.2412e-05 & 1.0564	 \\ [1ex] 
\hline
\end{tabular}
\caption{Properties of Novec 7000 computed or interpolated based on several references \cite{3MNovec,widiatmo2001equations,ohta2001liquid,rausch2015density,perkins2022measurement,aminian2022ideal} at the selected temperatures that are used in the experiment. Subscript v is for the vapor phase while subscript L is for the liquid phase.}
\label{table:NovecProp}
\end{sidewaystable}

Table \ref{table:NovecProp} summarizes the temperature-dependent properties of Novec 7000 for all ambient temperature settings that are used for the computation of $U_\text{sp}/C_\text{sound,L}$ using Eq. 6 in the main text. Although not being used in the computation of $U_\text{sp}/C_\text{sound,L}$, we provide here the dynamic viscosity $\mu_\text{v}$ and the specific heat capacity $C_\text{v}$ at constant pressure of vapor Novec 7000, the kinematic viscosity $\nu_\text{L}$ of liquid Novec 7000 and the specific heat ratio $\gamma$ as a reference. They are obtained by cubic interpolation of the data provided in \cite{rausch2015density} (see Table 4). 
The thermal conductivity of vapor Novec 7000 is provided in \cite{perkins2022measurement} with the empirical relationship:
\begin{equation}
    k_\text{v} = \frac{ A_0 + A_1\tau_\text{v} + A_2\tau_\text{v}^2 }{ a_0 + a_1\tau_\text{v} + a_2\tau_\text{v}^2 + a_3\tau_\text{v}^3 + a_4\tau_\text{v}^4 } 
\end{equation}
where $\tau_\text{v} = T_\text{v}/T_\text{c}$, $a_i$ are constants of values $a_0$ = 34.1702, $a_1$ = -49.3874, $a_2$ = 44.4355, $a_3$ = -11.058, $a_4$ = 1.0, and $A_0$ = 0.122535 W/m K, $A_1$ = -0.29099 W/m K and $A_2$ = 0.621481 W/m K are the dilute gas, residual and critical thermal conductivity, respectively. The thermal diffusivity of Novec 7000 vapor $\alpha_\text{v}$ can be calculated with Eq. \eqref{eq:alpha} using the properties of the vapor phase.

In general, at room temperature, the momentum (viscosity) and thermal (conductivity/diffusivity) transport properties of Novec 7000 in both liquid and vapor phases are slightly lower than that of water and air, but still of the same order of magnitude, except for the thermal conductivity of liquid Novec 7000 that is an order of magnitude lower than water. It is also good to note that the density ratio between the vapor and liquid $\rho_\text{v,0}/\rho_\text{L}$ for the equilibrium temperature settings in our experiment is always higher than the density ratio between air and water, which is around 0.0012.

\section{Justification of the heat-mass balance equation}

Here we briefly justify the appropriateness of using the mass balance equation (1) in the main document.

When a vapor pocket is entrapped below a dry disk, pressurization of the pocket may lead to condensation of vapor to restore the thermal equilibrium in the pocket. Note that when the disk is dry, condensation is likely to happen on the liquid vapor interface rather than on the disk, since spontaneous nucleation of droplets on the disk would require the pressure to rise from the vapor curve onto the spinodal curve, whereas vapor may directly condense onto the liquid vapor interface. 

Generally, the heat diffusivity $\alpha_\text{v}$ in the vapor phase is substantially larger than that in the liquid phase $\alpha_\text{L}$. For Novec 7000 in the range of temperatures used in our experiments, we have $\alpha_\text{v}/\alpha_\text{L} \approx 54-82$, which implies that the temperature of the vapor layer homogenizes much faster than that in the liquid, such that most of the condensation heat will flow into the liquid. In addition, even for substantial temperature changes $\Delta T$ in the order of $10$ K from the equilibrium value $T_0$, one can estimate the Jakob number as 
\begin{equation}
\mathit{Ja} = \frac{c_\text{p,v}\Delta T}{L} \approx 0.061-0.065\,,
\end{equation}  
where $c_\text{p,v}$ is the specific heat at constant pressure of the vapor phase and $L$ the latent heat of vaporization, such that the sensible heat changes may be neglected with respect to the latent heat. Consequently we may write in good approximation that the heat balance on the liquid-vapor interface is given by
\begin{equation}\label{eq:hm1}
-L \frac{dm_\text{v}}{dt} = S_\text{A} q''\,,
\end{equation}  
where $dm_\text{v}/dt$ is the time rate of change of the vapor mass in the pocket, $S_\text{A}$ its surface area, and $q''$ the heat flux away from the interface.

On the other hand, the heat conductivity $k_\text{v}$ into the vapor phase is much smaller than that in the liquid phase $k_\text{L}$, where for Novec 7000 $k_\text{v}/k_\text{L} \approx 0.14-0.16$, owing to the small vapor to liquid ratio density ($\rho_\text{v,0}/\rho_\text{L} \approx 0.002-0.004$), such that if we estimate the heat flux to the disk as $q''_\text{v} \approx k_\text{v} \Delta T/h_0$ and that into the liquid as $q''_\text{L} \approx k_\text{L} \Delta T/\sqrt{\pi\alpha_\text{L} t}$ (where $h_0 \approx 0.1$ mm is the thickness of the vapor pocket and $t \approx 1.0$ ms is the duration of the pocket pressurization, we find for the ratio
\begin{equation}
\frac{q''_\text{v}}{q''_\text{L}} \approx \frac{k_\text{v} \sqrt{\pi \alpha_\text{L} t}}{k_\text{L} h_0} \approx 0.017-0.019\,,
\end{equation}  
that is, the heat flux into the liquid is orders of magnitude larger than that into the vapor, such that we may take $q'' = q''_\text{L}$ in Eq.~\eqref{eq:hm1} and we arrive at Eq.~(1) in the main document:
\begin{equation}\label{eq:hm2}
-L \frac{dm_\text{v}}{dt} = q''_\text{L} = - S_\text{A} k_\text{L} \left. \frac{\partial T}{\partial n} \right|_{\text{interface}} \,,
\end{equation}   
with $\partial T/\partial n$ the normal component of the temperature gradient into the liquid.     

The vapor mass change in the bubble is then related to a change of the thermodynamic state in the vapor using Clausius-Clapeyron and ideal gas law. In doing so, we disregard that when (e.g.) vapor is disappearing from the pocket due to condensation, also some liquid is produced. More concretely we have
\begin{equation}\label{eq:mLiquid}
- \frac{\Delta m_\text{L}}{\Delta m_\text{v}} = \frac{\rho_\text{v,0}}{\rho_\text{L}}  \approx 0.002-0.004\,,
\end{equation}   
such that the liquid production may safely be neglected.    

\begin{figure*}
\includegraphics[width=\textwidth, trim=0cm 6cm 0cm 6cm]{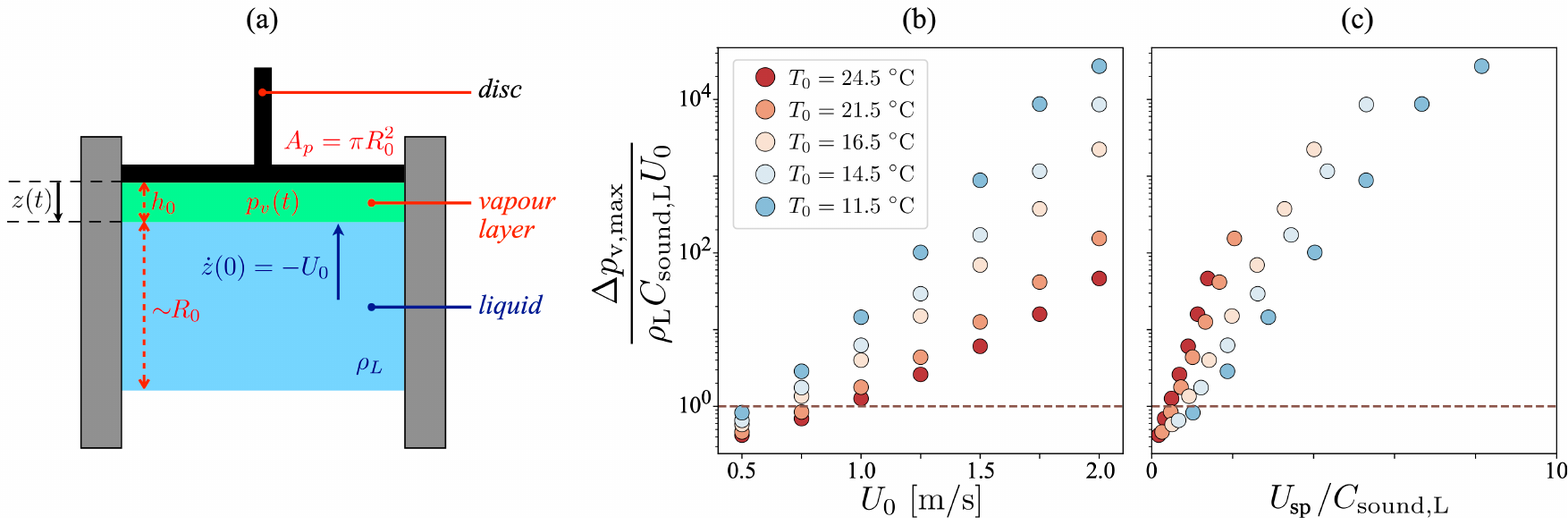}
\caption{\label{fig:Sup1} (a) Sketch of the Bagnold model for the disk impact. In this model the disk is at rest and the liquid is moving upwards (negative $z$-direction) with an initial velocity $\dot{z} = -U_0$, compressing a vapor layer of initial thickness $z(0) = h_0$. (b) Ratio of the maximum gauge pressure $\Delta p_\text{v,max}$ obtained from the Bagnold model and the water hammer pressure $\rho_\text{L}C_\text{sound,L}U_0$ on a logarithmic scale, as a function of the impact velocity $U_0$ for all the experimental settings used in the main document. (c) Same as in (b), but now plotted against the quantity $U_\text{sp}/C_\text{sound,L}$ introduced in Eq. (6) in the main document.}
\end{figure*}

\section{Justification of the water hammer pressure scale}

In this Section we provide some additional justification of using the water hammer pressure as the appropriate pressure scale. As argued in the main document, the collapse of the vapor pocket is very fast, such that it is expected that the water hammer pressure is the appropriate pressure scale. First of all, why is this the case? 

If the gauge pressure in the vapor pocket $\Delta p$ becomes large enough, the liquid will start to respond and added mass, the liquid mass that needs to be accelerated with the moving disk, will start to build up. This build up will typically happen at the speed of sound $C_\text{sound,L}$ of the liquid, such that one may write $m_A(t) = \rho_\text{L} A_p C_\text{sound,L} t$, where $A_p$ is the area of the disk. Assuming that the vapor pocket interface moves with the disk speed $U_0$, we then find that the  minimum pressure necessary for such a build up is given by Newton's second law   
\begin{equation}
\Delta p = \frac{F}{A_p} = \frac{d}{dt}\!\!\left[\frac{m_A(t)U_0}{A_p}\right] = \rho_\text{L} C_\text{sound,L} U_0\,,\label{eq:pV}
\end{equation}
i.e., it is equal to the water hammer pressure.

The second issue is that in writing down the water hammer pressure as the appropriate scale, we tacitly assume that pressures inside the vapor pocket may reach such high pressures. To give additional evidence that this is indeed the case, we here present a simple, one-dimensional model for the dynamics of the vapor pocket known as the Bagnold model \cite{bagnold1939interim,Ancellin2012,Brosset2013}, where we neglect phase change. Note that a semi-analytical model that extends the Bagnold's model to include phase change is proposed in \cite{Ancellin2012}.

The Bagnold model assumes a gas/vapor pocket of initial thickness $z(0) = h_0$ and pressure $p_\text{v,0}$ as sketched in Fig.~\ref{fig:Sup1}a, which is adiabatically pressurized by a liquid piston of known mass $m_p = c_A (4/3)\pi R_0^3\rho_\text{L}$, corresponding to the added mass of a circular disk of radius $R_0$ moving into a liquid, where $c_A$ is the so-called added mass coefficient \cite{brennenbook}. At $t=0$, the piston hits the pocket with a velocity $\dot{z}(0) = - U_0$, where $\dot{z}$ denotes the time derivative of $z(t)$ and with which one immediately writes down the initial value problem
\begin{subequations} 
\begin{align} 
m_p\ddot{z} &= A_p (p_\text{v}(t)-p_\text{v,0})\,,\label{eq:dim}\\
z(0) &= h_0\,\,;\,\,\,\,\dot{z}(0) = -U_0\,,\label{eq:dimBC}
\end{align} 
\end{subequations} 
for the time evolution of the gas/vapor pocket size $z(t)$. Here, $\dot{z}$ denotes the derivative of $z$ with respect to time, $A_p = \pi R_0^2$ is the surface area of the piston and $p_\text{v}(t)$ is related to $z(t)$ by the adiabatic compression law
\begin{equation}
p_\text{v}(t) z(t)^\gamma = p_\text{v,0} h_0^\gamma\,,\label{eq:pV}
\end{equation} 
where $\gamma$ is the specific heat ratio. Non-dimensionalising the equations using $\tilde{z} = z/h_0$, $\tilde{t} = U_0t/h_0$, and $\tilde{p}_\text{v} = p_\text{v}/p_\text{v,0}$ the initial value problem now becomes 
\begin{subequations} 
\begin{align} 
\ddot{\tilde{z}} &= S (\tilde{z}^{-\gamma}-1)\,,\label{eq:nondim} \\
\tilde{z}(0) &= 1\,\,;\,\,\,\,\dot{\tilde{z}}(0) = -1\,,\label{eq:nondimBC}
\end{align} 
\end{subequations} 
where $\ddot{\tilde{z}}$ now indicates the second derivative with respect to dimensionless time, and the impact parameter $S$ is given by
\begin{equation}\label{eq:S}
S = \frac{3}{4c_A}\frac{p_\text{v,0}}{\rho_\text{L}U_0^2}\frac{h_0}{R_0}\,,
\end{equation} 
and the pressure is given by
\begin{equation}
\tilde{p}_\text{v} = \tilde{z}^{-\gamma}\,.\label{eq:pVnondim}
\end{equation} 
It is straightforward to write down the first integral of Eq.~\eqref{eq:nondim}, multiplying with $d\tilde{z} = \dot{\tilde{z}}d\tilde{t}$ and integrating, as
\begin{equation}\label{eq:E}
E = \tfrac{1}{2}\dot{\tilde{z}}^2 + S \tilde{z}\left[1+\frac{\tilde{z}^{-\gamma}}{\gamma-1}\right] = \tfrac{1}{2}\dot{\tilde{z}}^2 + S \tilde{p}_\text{v}^{-1/\gamma}\left[1+\frac{\tilde{p}_\text{v}}{\gamma-1}\right]\,,
\end{equation} 
where $E$ is a constant of motion that we may obtain from the initial conditions~\eqref{eq:dimBC} as $E = 1/2 + S\gamma/(\gamma-1)$.  

When the pocket is maximally compressed, we have $\dot{z} = 0$, with minimal size ($\tilde{z}_{min}$) and maximal pressure ($\tilde{p}_\text{v,max}$), which with \eqref{eq:E} can now be expressed in terms of $S$ and $\gamma$ as
\begin{equation}\label{eq:pmax} 
S \tilde{p}_\text{v,max}^{-1/\gamma}\left[1+\frac{\tilde{p}_\text{v,max}}{\gamma-1}\right]= \tfrac{1}{2} + \frac{S\gamma}{\gamma-1}\,.
\end{equation}   
To obtain $\tilde{p}_\text{v,max}$ for the experiments presented in the main document, we compute $S$ using $p_\text{v,0}$, $\rho_\text{v,0}$, $\rho_\text{L}$, and $U_0$ from the experimental settings together with the relation $h_0/R_0 = \lambda\rho_\text{v,0}/\rho_\text{L}$ with $\lambda = 2.1$ also used in the main text. Note that we have used $c_A=1$, although values slightly smaller than 1 have been reported in the literature for the quite different situation of a disk fully immersed and moving through a liquid in the context of potential flow (see, e.g., \cite{brennenbook}). Subsequently, \eqref{eq:pmax} is solved for $\tilde{p}_\text{v,max}$ from which we then compute the dimensional $\Delta p_\text{v,max} = p_\text{v,0}(\tilde{p}_\text{v,max}-1)$. We subsequently divide $\Delta p_\text{v,max}$ by the water hammer pressure $\rho_\text{L}C_\text{sound,L}U_0$, and plot the result (on a logarithmic scale) as a function of $U_0$ (inset) in Fig.~\ref{fig:Sup1}b, and of the quantity $U_\text{sp}/C_\text{sound,L}$ introduced in the main document in Fig.~\ref{fig:Sup1}. Clearly, for all cases where condensation could start to play a role, the maximum pressure $\Delta p_\text{v,max}$ is much larger than the water hammer pressure $\rho_\text{L}C_\text{sound,L}U_0$. Only for small $U_0$ and higher $T_0$ the pressure in the pocket is not expected to reach the water hammer pressure.

If the pressure in the vapor pocket is always smaller than the water hammer pressure, added mass only starts building up gradually when the liquid comes into contact with the disk itself.  

Finally, it should be noted that $U_\text{sp}/C_\text{sound,L}$ obtained from Eq.~(6) in the main manuscript is almost an order of magnitude larger than the experimentally measured spreading speeds. This could be due to any number of approximations made in the derivation of the scaling result Eq.~(6). We have chosen not to add a numerical factor to rescale the computed $U_\text{sp}/C_\text{sound,L}$ towards the experimental one, and therefore this discrepancy needs to be taken into account when interpreting the above Fig.~\ref{fig:Sup1}c, or Fig.~4 in the main document.  

\begin{figure}
\includegraphics[width=0.8\linewidth]{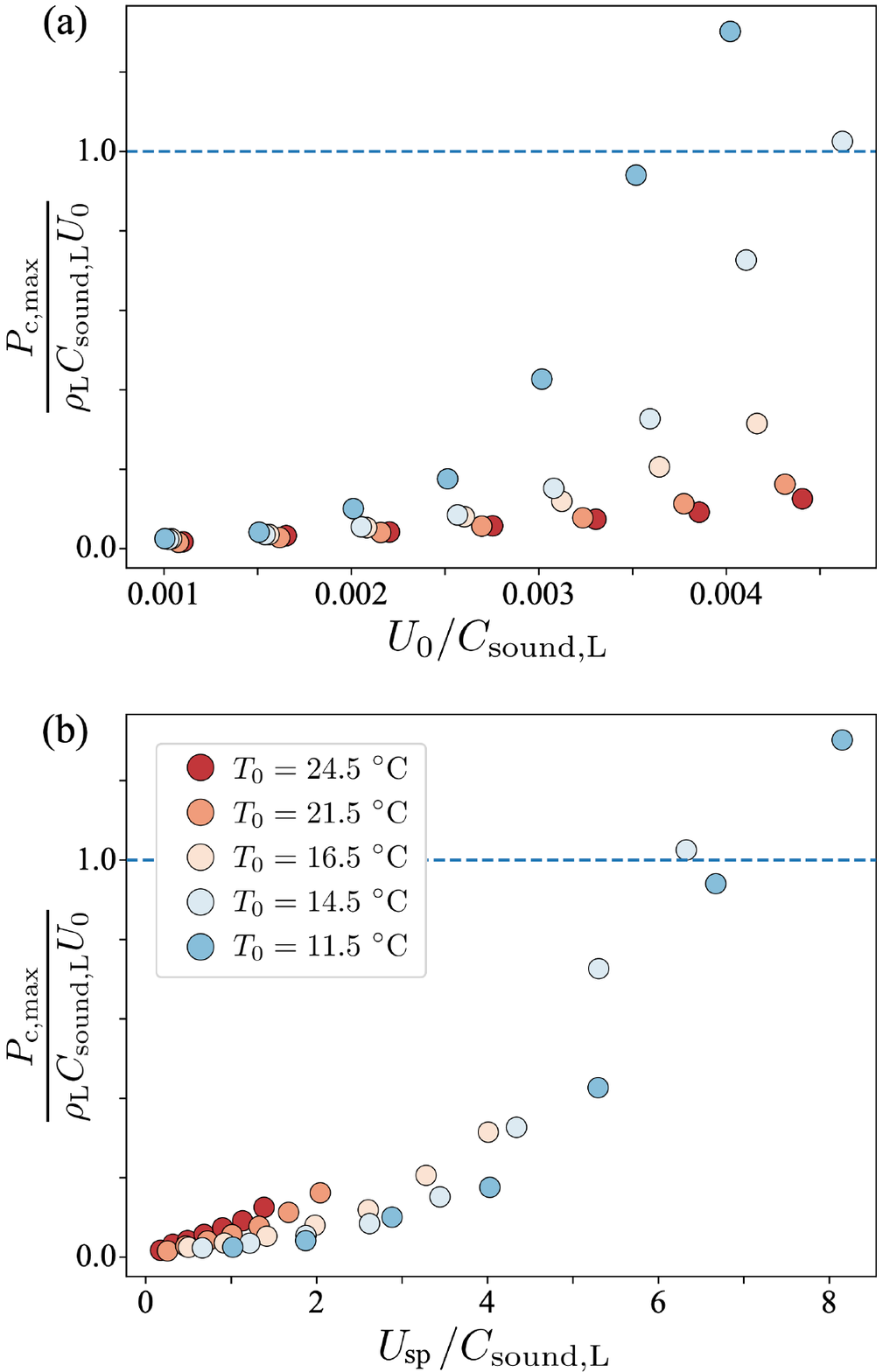}
\caption{\label{fig:Sup2}(a) Maximum central impact pressure $P_\text{c,max}$, rescaled
with the water hammer pressure scale $\rho_\text{L}C_\text{sound,L}U_0$, and plotted against the
impact Mach number $U_0/C_\text{sound,L}$ in the liquid at different ambient temperature $T_0$. (b) Rescaled maximum central impact pressure $P_\text{c,max}/(\rho_\text{L}C_\text{sound,L}U_0)$ versus the quantity $U_\text{sp}/C_\text{sound,L}$ defined in Eq. (6) in the main text. In both cases the limiting pressure from the Joukowsky equation ($P = \rho_\text{L}C_\text{sound,L}U_0$) is indicated by the horizontal black dashed line.}
\end{figure}

\section{Inertial vs. compressible rescaling\\of impact pressure}

In Figs. 2 and 4 of the main document we use the inertial pressure scale $\rho_\text{L}U_0^2$ to rescale the maximum pressure recorded in the central pressure sensor during impact. However, one could argue that, since the main line of reasoning leads to the compressible response of the liquid during the build-up of added mass being of key importance for the cavity compression that eventually leads to condensation, a more appropriate pressure scale may be that of the water hammer pressure $\rho_\text{L} C_\text{sound,L}U_0$ that is associated with compressible liquid response, at least for those cases where very high pressures are observed. To show that this is not the case, in Fig.~\ref{fig:Sup2}a we plot the same data $P_\text{c,max}$ as in Figs. 2 and 4 of the main document, but now rescaled with the water hammer pressure scale $\rho_\text{L} C_\text{sound,L}U_0$, as a function of the Mach number $\textrm{Ma} = U_0/C_\text{sound,L}$. First, it is clear that the measured central pressure $P_\text{c,max}$ does not appear to be limited by the water hammer pressure $\rho_\text{L} C_\text{sound,L}U_0$, as suggested by the Joukowsky equation for impact of compressible liquids. The measured $P_\text{c,max}/(\rho_\text{L} C_\text{sound,L}U_0)$ data crosses the value $1$ without any sign of converging to a horizontal asymptotic value. In fact, by careful alignment of the disk we have measured rescaled central pressure values as high as $3.9$ (for $T_0 = 11.5$ $^\circ$C and $U_0 = 2.0$ m/s). Secondly, it is noted that the Mach number is small, $\textrm{Ma} < 0.005$, for which dominance of compressible effects in the impact response is not expected. 

For completeness, in Fig.~\ref{fig:Sup2}b we plot the compressible analogue of Fig. 4 of the main document, where we plot the rescaled central maximum pressure $P_\text{c,max}/(\rho_\text{L} C_\text{sound,L}U_0)$, as a function of the quantity $U_\text{sp}/C_\text{sound,L}$ defined in Eq. (6) in the main text. Clearly, rescaling with the water hammer pressure $\rho_\text{L} C_\text{sound,L}U_0$ instead of the inertial pressure scale $\rho_\text{L} U_0^2$ completely fails to collapse the data.

All of the above together confirm that the observed large impact pressures cannot be solely due to a compressible liquid response and another mechanism, namely condensation, needs to be incorporated in the analysis of the entrapped air pocket dynamics and the observed anomalously large impact pressures. In addition, the observation that $P_\text{c,max}$ may considerably exceed the water hammer limit, suggests a very fast disappearance of the vapor from the entrapped pocket, possibly even leading to large suction pressures inside the vapor pocket. Clearly, this aspect needs further investigation.

\bibliographystyle{apsrev4-2}
\bibliography{smref}